\documentclass[3p,times]{elsarticle}

\usepackage{lineno,hyperref}

\usepackage{epsfig}
\usepackage{graphicx}
\usepackage{psfrag}
\usepackage{amsmath,amssymb}
\usepackage{colordvi}
\usepackage{amsfonts}
\usepackage{enumerate}
\usepackage{slashed}
\usepackage{color}
\usepackage{xcolor}

\def\bea#1\eea{\begin{align}#1\end{align}}

\begin{document}

%\preprint{JLAB-THY-15-2047}

\title{Scheme dependence and Transverse Momentum Distribution interpretation of
Collins-Soper-Sterman resummation}

\author[jlab]{Alexei Prokudin}
\ead{prokudin@jlab.org}

\author[lbnl]{Peng Sun}
\ead{psun@lbl.gov}

\author[lbnl]{Feng Yuan}
\ead{fyuan@lbl.gov}

\address[jlab]{Jefferson Lab, 
                   12000 Jefferson Avenue, 
                   Newport News, VA 23606, USA}

\address[lbnl]{Nuclear Science Division, 
                   Lawrence Berkeley National Laboratory, 
                   Berkeley, CA 94720, USA}

\begin{abstract}
Following an earlier derivation by Catani-de Florian-Grazzini (2000) on the scheme
dependence in the Collins-Soper-Sterman (CSS) resummation formalism in hard 
scattering processes, we investigate the scheme dependence of the Transverse 
Momentum Distributions (TMDs) and their applications.
By adopting a universal $C$-coefficient function associated with the integrated
parton distributions, the difference between various TMD schemes can be
attributed to a perturbative calculable function depending on the hard momentum
scale. We further apply several TMD schemes to the Drell-Yan
process of lepton pair production in hadronic collisions, and find that the constrained
non-perturbative form factors in different schemes are remarkably consistent with each other and with that of the 
standard CSS formalism. 
\end{abstract}

%\pacs{12.38.Bx, 12.39.St, 13.85.Hd, 13.88.+e}
\maketitle

\section{Introduction}
The Transverse Momentum Distributions (TMDs) and the nucleon tomography 
in momentum space have attracted strong interest in recent years~\cite{Boer:2011fh,Accardi:2012qut}.  
TMDs provide
a unique opportunity to investigate the novel correlations between
the parton momentum and the nucleon spin. They unveil the strong interaction
QCD dynamics in a manifest way, such as the gauge invariance
leading to the sign change \cite{Collins:2002kn,Brodsky:2002cx} of certain TMDs in different hard scattering
processes, and the QCD factorization and
evolution which are crucial for predicting the scale dependence of the spin asymmetries.
On the theory side, the TMDs are not straightforward extensions~\cite{Collins:2003fm} of the conventional 
 collinear parton distribution
functions (PDFs). They hold special properties 
 that differ them from collinear PDFs 
 and play important
roles in high energy scattering.
The associated phenomena are direct consequences of perturbation gauge theory 
 computation of  the famous Sudakov form factors~\cite{Sudakov:1954sw} back in 1950s. 

When one applies the TMD factorization to  physical processes, one has to
consider the associated QCD dynamics in the definition of TMDs and in the phenomenological
studies. Especially, large logarithmic corrections from high order perturbative calculations 
have to be taken into account and resummed~\cite{Collins:1984kg} to all orders. 
In addition, the naive gauge invariant TMD definition contains the so-called
light-cone singularities at higher orders and needs to be regulated~\cite{Collins:1981uk}. 
Several 
ways to implement such a regularization 
have been proposed in the literature and they introduce the scheme dependence
in  TMDs and their applications~\cite{Collins:1981uk,Collins:2011zzd,Ji:2004wu,Ji:2004xq,GarciaEchevarria:2011rb,Ji:2014hxa}. 
The goal of this paper is to investigate such scheme dependence, 
which is of crucial importance for applying the TMDs
in hard scattering processes and extracting the associated nucleon structure 
from  experiments.

In the context of the standard Collins-Soper-Sterman (CSS) resummation
formalism~\cite{Collins:1984kg}, the TMDs are expressed in terms of the collinear parton distributions via
an additional factorization at small $b \ll 1/\Lambda_{QCD}$, where $b$ represents the Fourier conjugate
variable associated with the transverse momentum $k_\perp$. 
The final expressions for the measured  cross-sections differential in transverse momentum of the observed particles 
do not depend how we define the TMDs at the first place when such relations to collinear PDFs are used.
In other words, in all  TMD 
formalisms of Refs.~\cite{Collins:1981uk,Collins:2011zzd,Ji:2004wu,Ji:2004xq,GarciaEchevarria:2011rb,Ji:2014hxa},
one will 
obtain the same results as that of  the standard CSS resummation.
 However, as discussed in an early paper by Catani, de Florian, and 
Grazzini~\cite{Catani:2000vq}, 
even in conventional CSS formalism 
there is freedom to separate  
the so-called
 hard factor, $H$,  
which depends on the running coupling at the hard momentum scale $Q$
from the $C$-coefficient functions associated with the integrated parton 
distribution functions where running coupling depends on $\mu_b=c_0/b$ with $c_0=2e^{-\gamma_E}$. 
It was  referred to in Ref.~\cite{Catani:2000vq} as the scheme 
dependence of CSS resummation. The relation between
different schemes was further demonstrated by an order by order proof~\cite{Catani:2000vq}. 
The relevant derivations with explicit results up to next-to-next-leading order
for Drell-Yan, Higgs boson, di-photon production processes have
been extensively discussed in Ref.~\cite{Catani:2013tia}.
The same argument  applies to the scheme dependence in the TMD formalism as well~\cite{Collins:1984kg,
Collins:1981uk,Collins:2011zzd,Ji:2004wu,Ji:2004xq,GarciaEchevarria:2011rb,Ji:2014hxa}.
By adopting a universal $C$-coefficient function associated with the collinear 
parton distributions~\cite{Catani:2000vq}, the connections between different 
schemes can be attributed to the 
hard coefficients 
and can be established order by order in perturbation 
theory. As a result, all the TMD scheme 
dependence can be accounted for  and the schemes can be unified and compared to the 
standard CSS resummation 
in description of  the experimental data in 
phenomenological studies.

Furthermore, this 
unification
 provides an attractive 
 interpretation for the CSS resummation,
from which we have a clear TMD interpretation 
of the hard scattering processes.
To establish this, we apply this scheme in the global analysis of the Drell-Yan 
process of
lepton pair production in $pp$ collisions, and fit the associated non-perturbative form
factors. 
In the calculations, we adopt the so-called $b_*$-prescription
and derive the relevant perturbative coefficients following 
the procedure of Ref.~\cite{Catani:2000vq}.
Our results show that the non-perturbative form factors are remarkably 
consistent with that in
the standard CSS scheme. This is a very important
result that will clarify the confusing issues of 
the TMD evolution which has been intensively explored in recent years.

The rest of the paper is organized as follows. In Sec.~\ref{TMD Schemes}, we briefly introduce
  the TMD schemes in hard scattering processes, and derive the relevant
coefficients. In Sec.~\ref{NP factors}, we fit  the experimental data of Drell-Yan type of
hard processes in hadronic collisions and constrain the associated non-perturbative
form factors. And finally, we conclude our paper in Sec.~\ref{Conclusions}.

\section{TMD Schemes \label{TMD Schemes}}

Let us start with the standard CSS
resummation formalism for Drell-Yan lepton pair production
processes at low transverse momentum:
$A (P_A)+B(P_B) \to \gamma^* (q) +X \to \ell^+ + \ell^ -
+X$, where $P_A$ and $P_B$ represent the momenta of the incoming 
hadrons $A$ and $B$, respectively.
The differential cross section can be written as~\cite{Collins:1984kg},
\begin{eqnarray}
\frac{d^4\sigma}{dydQ^2d^2q_\perp}&=&
\sigma_0^{\rm (DY)}\left[\int \frac{d^2b}{(2\pi)^{2}} \, e^{i \vec{q}_\perp\cdot \vec{b}}\, \widetilde
{W}_{UU}(Q;b) +Y_{UU}(Q;q_\perp) \right]\ ,
\end{eqnarray}
where $q_\perp$ and $y$ are transverse momentum and rapidity of the lepton pair, respectively,
$\sigma_0^{\rm (DY)}=4\pi\alpha_{em}^2/3N_csQ^2$ with $s=(P_A+P_B)^2$.
In the above equation, the first term is dominant in the $q_\perp\ll Q$ region and $\widetilde{W}_{UU}$ denotes the all-order
resummation result which has the following form~\cite{Collins:1984kg,Catani:2000vq}:
\begin{eqnarray}
\widetilde {W}_{UU}(Q;b)&=&H^{(DY)}(\alpha_s(Q))\, e^{-{ S}(Q^2,b)}\,  
\sum_{q=q,\bar q} e_q^2 \, C_{q\gets i}^{(DY)}\otimes f_{i/A}(x_1,\mu_b)\, C_{\bar q\gets j}^{(DY)}\otimes f_{j/B}(x_2,\mu_b) \ ,\label{css}
\end{eqnarray}
where $\mu_b=c_0/b_*$ with $c_0=2e^{-\gamma_E}$ and $\gamma_E$ the Euler constant,
$x_{1,2}=Qe^{\pm y}/\sqrt{s}$ represent the momentum fractions carried by the incoming
quark and antiquark in the Drell-Yan processes, the symbol $\otimes$ for convolution
in $x_1$($x_2$) 
 and $f_{i/A}(x,\mu_b)$ and $f_{j/B}(x,\mu_b)$ stand
for the collinear integrated parton distribution functions at the scale $\mu_b$. 
 In Eq.~(\ref{css}),
$b_*$-prescription, $b \to b_*=b/\sqrt{1+b^2/b_{max}^2}$,
is introduced~\cite{Collins:1984kg}. The form factor ${S}(Q,b)$ contains
perturbative and nonperturbative parts, such that the total form factor for quarks can be written as ${S}(Q,b)={ S}_{pert}(Q,b_*)+{ S}_{NP}(Q,b)$,
\bea
S_{\rm pert}(Q,b)=\int_{\mu_b^2}^{Q^2}\frac{d\bar\mu^2}{\bar\mu^2}\left[A(\alpha_s(\bar\mu))\ln\frac{Q^2}{\bar\mu^2}+B(\alpha_s(\bar\mu))\right] \ ,
\label{spert}
\eea
where $A$, $B$ and $C$ coefficients calculable order by order in perturbation theory 
perturbative series $A=\sum_{n=1}^\infty A^{(n)} \left(\alpha_s/\pi\right)^n$, $B=\sum_{n=1}^\infty B^{(n)} \left(\alpha_s/\pi\right)^n$,
$C=\sum_{n=1}^\infty C^{(n)} \left(\alpha_s/\pi\right)^n$.
 The $A$, $B$, $C$ coefficients can be derived~\cite{Catani:2000vq},
\begin{eqnarray}
&&A_{\rm CSS}^{(1)}=C_F, ~~B_{\rm CSS}^{(1)}=-\frac{3}{2} C_F,
~~C_{\rm CSS}^{(1)}=\frac{C_F}{2}\left[(1-x)+\delta(1-x)\frac{\pi^2-8}{2}\right]\ ,\nonumber\\
&&A_{\rm CSS}^{(2)}= \frac{C_F}{2}\left(C_A\left(\frac{67}{18}-\frac{\pi^2}{6}\right)-\frac{5}{9}N_f\right)\ ,\nonumber\\
&&B_{\rm CSS}^{(2)}=C_F^2\left(\frac{\pi^2}{4}-\frac{3}{16}-3\zeta_3\right)
+C_FC_A\left(\frac{11}{36}\pi^2-\frac{193}{48}+\frac{3}{2}\zeta_3\right)
+C_FN_f\left(\frac{17}{24}-\frac{\pi^2}{18}\right) \ , \label{standard}
\end{eqnarray}
in the standard CSS scheme. 
In the standard CSS formalism, the hard coefficient $H_{\rm CSS}(\alpha_s(Q))\equiv 1$ for all orders. 

We would like to emphasize that the resummation
formula and the associated coefficients are uniquely determined, once the
scheme is fixed.
The reason is simple. In the perturbative calculations of hard
processes at low transverse momentum, the large logarithms depend on 
two separate scales: $Q$ and $1/b$, the hard momentum and the Fourier
conjugate of the traverse momentum $q_\perp$, respectively. 
The resummation of these large logarithms has to take the form
as in Eq.~(\ref{spert}), as a consequence of perturbation gauge
theory computation of Sudakov form factors~\cite{Sudakov:1954sw,Korchemsky:1987wg,Collins:1981uk}. Additional
factors in the CSS resummation come from the fact that the
collinear gluon splitting is proportional to $1/q_\perp^2$ (again a result of a gauge theory computation), for which the Fourier transformation leads
to $\ln (\mu_b/\mu)$ where $\mu$ represents the PDF scale. 
Therefore, the integrated parton distribution
is calculated at $\mu_b$ for canonical choice of the resummation. By
doing so, we also resum the logarithms associated with collinear
gluon radiation. The coefficients $A$, $B$, and $C$ can be obtained
from the factorization derivation, or by comparing to the fixed order
perturbative calculations. For phenomenological applications,
the CSS formalism has been very successful in Drell-Yan lepton 
pair production, $W^\pm$/$Z$ boson production in hadron collisions~\cite{Landry:2002ix,Qiu:2000ga,Qiu:2000hf,Kulesza:2002rh,Kulesza:2003wn,
Catani:2003zt,Bozzi:2003jy,Bozzi:2005wk,Bozzi:2007pn,Bozzi:2008bb,Bozzi:2010xn,Sun:2013hua,Su:2014wpa}.

As discussed in Ref.~\cite{Catani:2000vq}, there is a freedom  
to absorb $\alpha_s(Q)$ corrections from higher orders in the definition of hard coefficient  $H^{(DY)}(\alpha_s(Q))$ of Eq.~(\ref{css}). 
Then, the associated
$B$ and $C$ coefficients will be modified according to the renormalization group equations. 
This was referred to as the scheme dependence in the CSS resummation in Ref.~\cite{Catani:2000vq}. 
In the following, we will apply this idea to discuss the TMD interpretation of the CSS resummation formalism,
where the scheme dependence is essential in the TMD definition and factorization.

TMD factorization~\cite{Collins:2011zzd} aims at separating well defined TMD distributions in 
Eq.~(\ref{css}), such that the TMD distributions can be used in different processes in a universal manner.
The  $b$-space expression of $\widetilde{W}_{UU}$ in Eq.~(\ref{css}) thus can be rewritten in the TMD factorization in terms of a product of process independent TMDs and a process dependent hard factor:
\begin{equation}
\widetilde{W}_{UU}(Q;b) ={\cal H}^{\rm TMD}_{(DY)}(Q;\mu) \, \sum_{q=q,\bar q} e_q^2\,\tilde{f}_{q/A}^{(sub)}(x_1,b;Q,\mu) \, \tilde{f}_{\bar q/B}^{(sub)}(x_2,b;Q,\mu)
\ ,\label{tmd}
\end{equation}
where both the subtracted  quark distribution $\tilde f_q^{(sub)}$ and hard
factor ${\cal H}^{\rm TMD}_{(DY)}$ depend on the scheme we choose to regulate the
light-cone singularity in the TMD definition. In this paper we consider
three TMD schemes: (1) Ji-Ma-Yuan 2004~\cite{Ji:2004wu,Ji:2004xq}; (2) Collins 2011~\cite{Collins:2011zzd}; 
(3) Lattice~\cite{Ji:2014hxa} or Collins-Soper 1981~\cite{Collins:1981uk}. The 
so-called EIS scheme was shown to be  equivalent to Collins 2011 scheme~\cite{Collins:2012uy}. 
Moreover, because of usage of space-like gauge link in the lattice scheme, the results in this scheme coincide with the the original
Collins-Soper 81 scheme. Extensions
to other formalisms can follow accordingly. 

We take the example of Ji-Ma-Yuan 2004 scheme (JMY)~\cite{Ji:2004wu,Ji:2004xq}, where 
the unpolarized quark distribution is defined as
\begin{eqnarray}
f_q(x,k_\perp;\zeta,\mu_F,\rho)&=&\frac{1}{2}\int
        \frac{d\xi^-d^2b}{(2\pi)^3}e^{-ix\xi^-P^++i\vec{b}\cdot
        \vec{k}_\perp}  \left\langle
PS\left|\overline\psi(\xi^-,0,\vec{b}){\cal L}_{v}^\dagger(-\infty;\xi)\gamma^+{\cal L}_{v}(-\infty;0)
        \psi(0)\right|PS\right\rangle\ ,\label{tmdun}
\end{eqnarray}
with the gauge link $ {\cal L}_{v}(-\infty;\xi) \equiv \exp\left(-ig\int^{-\infty}_0 d\lambda
\, v\cdot A(\lambda v +\xi)\right)$. The above definition contains the
light-cone singularity if we take the gauge link along the light-front direction, $v^2=0$. 
The way to regulate this singularity and subtract soft gluon contribution defines the scheme for TMDs. 
 In the JMY scheme, the gauge link is chosen to be slightly off-light-cone, 
 such that 
$n=(1^-,0^+,0_\perp)\to v=(v^-,v^+,0_\perp)$
with $v^-\gg v^+$. Similarly, for the TMD antiquark distribution, $\bar v$
was introduced, $\bar v=(\bar v^-,\bar v^+,0_\perp)$ with $\bar v^+\gg \bar v^-$.
Because of the additional $v$ and $\bar v$, there are additional invariants:
$\zeta_1^2=(2v\cdot P_A)^2/v^2$, $\zeta_2^2=(2\bar{v}\cdot P_B)^2/\bar{v}^2$,
and $\rho^2=(2v\cdot \bar v)^2/v^2\bar{v}^2$. Accordingly, the soft factor
is defined as,
\begin{equation}
S^{v,\bar v}(b)={\langle 0|{\cal L}_{\bar
v}^\dagger(b_\perp) {\cal
L}_{v}^\dagger(b_\perp){\cal L}_{v}(0){\cal
L}_{\bar v}(0)  |0\rangle   }\, . \label{softg}
\end{equation}
Following the subtraction procedure of Ref.~\cite{Collins:2011zzd}, we can define the subtracted TMDs in
$b$-space in the JMY scheme as,
\begin{eqnarray}
\tilde{f}_{q(JMY)}^{(sub)}(x,b;\zeta,\mu_F,\rho)&=& \frac{\tilde{f}_q(x,b;\zeta,\mu_F,\rho)}{\sqrt{S(b;\rho,\mu_F)}}\ ,
\end{eqnarray}
where $\tilde{f}_q(x,b;\zeta,\mu_F,\rho)$ is the $b$-space expression for the un-subtracted
TMD of Eq.~(\ref{tmdun}).
The evolution equations are derived for the TMDs: one is the energy evolution equation respect to $\zeta$, 
the so-called Collins-Soper evolution equation~\cite{Collins:1981uk} and the renormalization 
group equation associated with the factorization scale $\mu_F$ and related to 
anomalous dimensions of the distribution $\tilde{f}$. 
After solving the evolution
equations and expressing the TMDs in terms of the integrated parton
distributions to have a complete resummation results, we can write,
\begin{eqnarray}
\tilde{f}_{q(JMY)}^{(sub)}(x,b;\zeta^2=\rho Q^2,\mu_F=Q, \rho)&=& e^{-{ {S}_{pert}^q(Q,b_*)}-{S}_{\rm NP}^q(Q,b)}
\, \,\widetilde{{\cal F}}_q^{\rm JMY}\left(\alpha_s(Q);\rho\right)\nonumber\\
 &&\times \sum_i\, C_{q\gets i}\otimes f_i(x,\mu_b)\ , \label{tmdqf}
\end{eqnarray}
where we have chosen the energy parameter $\zeta^2=\rho Q^2$ and the factorization 
scale $\mu_F=Q$ to resum large logarithms~\cite{Ji:2004wu,Ji:2004xq}. The perturbative form factor
${S}_{pert}^q$ contains contributions from the Collins-Soper evolution kernel
and the renormalization equation respect to the factorization scale $\mu_F$ 
 and $\mu_b$. 
Similar to the CSS resummation, $b_*$-prescription
was applied.
 In the above equation, we have also followed the derivations of Ref.~\cite{Catani:2000vq} to
include the $\rho$-dependence in the hard factor $\widetilde{\cal F}_q$ by applying  the renormalization group equation of 
running coupling $\alpha_s$. By doing that, the $C$-coefficients are much simplified and have the following  universal TMD form~\footnote{In principle, we can also choose $C_{\rm CSS}$ of 
Eq.~(\ref{standard}) for the
$C$-coefficients, which will go back to the standard CSS resummation for phenomenological
applications. We chose these coefficients for simplicity.},
 \bea
C_{q\gets q'}^{(TMD)}(x,\mu_b) &=\delta_{q'q} \left[\delta(1-x)+\frac{\alpha_s}{\pi}\left(\frac{C_F}{2}(1-x) \right) \right]\; , \label{univerc}
\\
C_{q\gets g}^{(TMD)}(x,\mu_b) &= \frac{\alpha_s}{\pi} {T_R} \, x (1-x)\; ,  \label{eq:cf1}
\eea
for the quark-quark and quark-gluon splitting case. A universal $C$-function
in the CSS resummation formalism has also been emphasized in Ref.~\cite{Catani:2013tia}. From
the results in Ref.~\cite{Ji:2004wu}, see, for example, Eq.~(36) of \cite{Ji:2004wu}, 
we obtain
\begin{equation}
\widetilde{\cal F}_q^{\rm JMY}\left(\alpha_s(Q);\rho\right)=1+\frac{\alpha_s}{2\pi}C_F\left(\ln\rho-\frac{\ln^2\rho}{2}-\frac{\pi^2}{2}-2\right) \ .\label{jmyf}
\end{equation}
The above equations are derived based on the perturbative calculation and 
the associated QCD factorization for the TMDs. They apply to all  TMD 
schemes~\cite{Collins:1981uk,Collins:2011zzd,Ji:2004wu,Ji:2004xq,GarciaEchevarria:2011rb,Ji:2014hxa}
mentioned above.
The collinear divergence in the TMDs can be factorized into the integrated parton 
distributions as shown in Eq.~(\ref{tmdqf}) at small $b \ll 1/\Lambda_{QCD}$. 
For large $b$, a non-perturbative function has to be included. 
The universal $C$-coefficient function is adopted to simplify the final expression for the TMDs
and minimize the higher order corrections associated with the integrated parton distributions.

Similarly, for the Collins 2011 (JCC) scheme, we have~\cite{Collins:2011zzd,Aybat:2011zv},
\begin{eqnarray}
\tilde{f}_{q(JCC)}^{(sub)}(x,b;\zeta_c^2=Q^2,\mu_F=Q)&=& e^{-{ {S}_{pert}^q(Q,b_*)}-{S}_{\rm NP}^q(Q,b)}
\, \, \widetilde{{\cal F}}_q^{\rm JCC}\left(\alpha_s(Q)\right)\nonumber\\
&&\times \sum_i \, C_{q\gets i}^{(TMD)}\otimes f_i(x,\mu_b)\ , \label{tmdqfc}\\
 \widetilde{\cal F}_q^{\rm JCC}\left(\alpha_s(Q)\right)&=&1+{\cal O}(\alpha_s^2) \ ,\label{jccf}
\end{eqnarray}
where $\zeta_c$ is the regulation parameter in JCC scheme and the
$\alpha_s$ correction in $\widetilde{\cal F}_q^{\rm JCC}$  vanishes~\footnote{There is an ambiguity for the
ultra-violet (UV) subtraction:
 an additional term of $\pi^2/12$ should be added
in $\alpha_s$ correction if we follow the standard $\overline{\rm MS}$ subtraction used in the standard CSS. Here we adopt
Collins-11 prescription for the UV subtraction.}. Again, we emphasize
that $C$-coefficient takes the same form as that in Eq.~(\ref{univerc}). 
Therefore, the scheme dependence in the TMDs only comes
from the hard function $\widetilde{\cal F}_q$ as we have shown in the above 
equation.

Recently, there has been a motivated study to formulate the TMDs on 
lattice, where a different subtraction scheme was adopted, for which we
have~\cite{Ji:2014hxa}
\begin{eqnarray}
\tilde{f}_{q(Lat.)}^{(sub)}(x,b;\zeta^2=Q^2,\mu_F=Q)&=& e^{-{ {S}_{pert}^q(Q,b_*)}-{S}_{\rm NP}^q(Q,b)}
\, \, \widetilde{{\cal F}}_q^{\rm Lat.}\left(\alpha_s(Q)\right)\nonumber\\
&&\times \sum_i\, C_{q\gets i}^{(TMD)}\otimes f_i(x,\mu_b)\ , \label{tmdqflat}\\
\widetilde{\cal F}_q^{\rm Lat.}\left(\alpha_s(Q)\right)&=&1+\frac{\alpha_s}{2\pi}C_F\left(-2\right) \ ,\label{latf}
\end{eqnarray}
where the regulator $\zeta$ is defined as $\zeta^2=(2n_z\cdot P)^2/(-n_z^2)$ with space-like $n_z$:
$n_z^2=-1$, $n_z\cdot P=-P_z$.
As we mentioned above, lattice scheme uses  the same space-like gauge link as the original Collins-Soper 1981 scheme,
that is why the final expression for TMD are the same in both schemes.

Applying the above TMDs into the factorization formula of Eq.~(\ref{tmd}), 
and comparing to that in Eq.~(\ref{css}), we will find that the TMDs actually
provide a special scheme for the CSS resummation in the
context of Ref.~\cite{Catani:2000vq}.  We can immediately
derive the relevant coefficients,
\begin{equation}
H_{\rm TMD}^{(DY)}\left(\alpha_s(Q)\right)=\widetilde{\cal F}_q\left(\alpha_s(Q)\right)\times\widetilde{\cal F}_{\bar q}\left(\alpha_s(Q)\right)\times 
{\cal H}^{\rm TMD}_{(DY)}\left(Q;Q\right)\ , \label{eq:htmd}
\end{equation}
which will enter into Eq.~(\ref{css}) for phenomenological applications.
Because the $C$-coefficients are universal among different TMD schemes,
we conclude that $H_{\rm TMD}^{(DY)}$ will be the same in all of the three 
schemes discussed above. In particular, in the JMY scheme, 
all three factors in Eq.~\eqref{eq:htmd} depend on $\rho$, however 
the final result for $H_{\rm TMD}^{(DY)}$ does not depend on $\rho$.
This demonstrates that all the TMD factorization schemes are 
equivalent in the context of the CSS resummation formalism, which
will be used in the phenomenological applications. This can be 
verified from the above explicit results and from the associated hard factors
calculated for different schemes at the one-loop order, and order by order 
proof can be done accordingly.

Further comparison also indicates that the perturbative and non-perturbative 
form factors for the quark and antiquark can be related to that in the CSS
formalism Eq.~(\ref{css}),
\begin{eqnarray}
&S_{pert}^q(Q,b_*)=S_{pert}^{\bar q}(Q,b_*)=S_{perp}(Q,b_*)/2 \ ,\label{spertq}\\
&S_{\rm NP}^q(Q,b)+S_{\rm NP}^{\bar q}(Q,b)=S_{\rm NP} (Q,b)\ ,\label{snpq}
\end{eqnarray}
where the perturbative form factor $S_{pert}(Q,b_*)$ takes the form of Eq.~(\ref{spert})
with $A$ and $B$  coefficients for a particular TMD scheme.
The above equation for the perturbative form factors can be verified explicitly 
from one-loop results in the TMD factorization of Refs.~\cite{Collins:2011zzd,Ji:2004wu,Ji:2004xq,Ji:2014hxa}. 
Higher orders can be calculated in perturbative expansion. 

From the above one-loop results for $\widetilde{\cal F}_{q,\bar q}$ and the relevant
hard factors in the TMD factorization calculated in Refs.~\cite{Ji:2004wu,Ji:2004xq,Collins:2011zzd,Ji:2014hxa},
\bea
{\cal H}^{\rm TMD\, JCC}_{(DY)}(Q;\mu) &= 1 +  \frac{\alpha_s(\mu)}{2\pi}C_F\left( 3 \ln \frac{Q^2}{\mu^2} - \ln^2  \frac{Q^2}{\mu^2} + \pi^2 -8\right) \, ,\\
{\cal H}^{\rm TMD\, JMY}_{(DY)}(Q;\mu) &= 1 +  \frac{\alpha_s(\mu)}{2\pi}C_F\left( (1+\ln \rho^2) \ln \frac{Q^2}{\mu^2} - \ln^2  \rho + \ln^2 \rho +2 \pi^2 -4\right) \, ,\\
{\cal H}^{\rm TMD\, Lat.}_{(DY)}(Q;\mu) &= 1 +  \frac{\alpha_s(\mu)}{2\pi}C_F\left( \ln \frac{Q^2}{\mu^2} + \pi^2 -4\right)\, ,
\eea
 we 
obtain the one-loop expression for $H_{\rm TMD}$ as,
\begin{eqnarray}
H_{\rm TMD}^{(1){(DY)}}&=&\frac{1}{2}C_F\left(\pi^2-8\right)\ , \label{tmdh}
\end{eqnarray}
so that 
\begin{eqnarray}
H_{\rm TMD}^{(DY)}(Q)&=&1 + \frac{\alpha_s(Q)}{2 \pi}C_F\left(\pi^2-8\right)\ , \label{tmdh_factor}
\end{eqnarray}

For $B$ and $C$ coefficients, following the derivation of Ref.~\cite{Catani:2000vq}, we will obtain 
\begin{eqnarray}
C_{\rm TMD}^{(1)}&=&C^{(1)}_{\rm CSS}-\delta(1-x)\, H_{\rm TMD}^{(1){(DY)}}/2\ ,\nonumber\\
B_{\rm TMD}^{(2)}&=&B^{(2)}_{\rm CSS}- \beta_0\, H_{\rm TMD}^{(1){(DY)}}\ ,\label{tmdbc}
\end{eqnarray}
where $\beta_0=\frac{11}{12}C_A-\frac{N_f}{6}$, and $A^{(1,2)}$ and $B^{(1)}$ 
remain the same as the standard CSS scheme. We will apply these coefficients in the
next section to analyze the Drell-Yan type of lepton pair production in 
hadronic processes to constrain the associated non-perturbative form factors.

Following the arguments of Ref.~\cite{Catani:2000vq}, the process-dependence 
is included in $H$ in Eq.~(\ref{css}), so that $C$-coefficients
will be universal. 
% %AP In the sense, 
We can apply the same $C$-functions to the 
quark distributions in other processes, such as the Semi-Inclusive Deep Inelastic
Scattering (SIDIS), for which we have the standard CSS resummation coefficient
\bea
B_{CSS}^{(2)(SIDIS)}=B_{CSS}^{(2)}-\beta_0\, C_F\, \pi^2/2 \, ,
\eea
with $B_{CSS}^{(2)}$ from Eq.~(\ref{standard}) and all other coefficients that have
been listed in Refs.~\cite{Nadolsky:1999kb,Nadolsky:2000ky,Sun:2013hua}. The hard function for SIDIS is
\begin{eqnarray}
H_{\rm TMD}^{(SIDIS)}(Q)&=&1 + \frac{\alpha_s(Q)}{2 \pi}C_F\left(-8\right)\ , \label{tmdh_factor_sidis}
\end{eqnarray}
if we choose the TMD scheme for this process.

\section{Non-perturbative Form Factors and TMD Interpretation \label{NP factors}}

As we mentioned above, we will apply the $b_*$-prescription for the 
non-perturbative form factors. We will follow the SIYY parameterization~\cite{Su:2014wpa}.
This parameterization is motivated by a phenomenological study~\cite{Sun:2013hua} and 
is inspired by  matching to perturbative calculations of the
Sudakov form factors~\cite{Collins:1985xx,Collins:2014jpa}.
It has the following form,
\begin{eqnarray}
{S}_{\rm NP}(Q, b) &=&  {g_2}\ln\left({b}/{b_*}\right)\ln\left({Q}/{Q_0}\right)+{g_1}b^2\; , \label{eq:siyy}
\end{eqnarray}
with the initial scale $Q_0^2=2.4$ GeV$^2$ and cut-off parameter $b_{max}=1.5$ GeV$^{-1}$.
The parameters $g_{1,2}$ have been fitted to the experimental data of Drell-Yan type   processes 
in Ref.~\cite{Su:2014wpa} using  the standard CSS formalism ($H\equiv 1$).
In the study of Ref.~\cite{Su:2014wpa} it was found that the experimental data are consistent with 
$x$-independent non perturbative factors. In the following 
studies, we will take the above simple form of Eq.~\eqref{eq:siyy}.  Since we will compare the TMD schemes
to the standard CSS scheme, we will keep all 
relevant parameters in non-perturbative factors fixed except for the
 changes in the coefficients $H^{(1)}$, $C^{(1)}$ and $B^{(2)}$.

We compare our results to the same set of the experimental data sets as those used in Ref.~\cite{Su:2014wpa}. The data sets 
include the Drell-Yan lepton pair production from fixed
target hadronic collisions from R209, E288 and E605~\cite{Ito:1980ev,Antreasyan:1981uv,Moreno:1990sf},
and $Z$ boson production in hadronic collisions from Tevatron
Run I and Run II~\cite{Affolder:1999jh,Abbott:1999wk,Abazov:2007ac,Aaltonen:2012fi}.
We proceed with the fit of the experimental data using  standard CSS, done in 
Ref.~\cite{Su:2014wpa} and the TMD-schemes described in this paper.
Notice that all TMD-schemes have exactly the same hard factor $H_{\rm TMD}^{(DY)}$, so by doing a single fit we effectively obtain underlying TMDs in either Collins 2011~\cite{Collins:2011zzd}, Lattice~\cite{Ji:2014hxa}, or Ji-Ma-Yuan 2004~\cite{Ji:2004wu,Ji:2004xq} schemes.
The fitted parameters are found to be,
\bea
\text{SIYY~\cite{Su:2014wpa}}&:~~g_1=0.212,~~~g_2=0.84\ , ~~\text{ total}~~\chi^2=168 ,\\
\text{SIYY}_\text{TMD}&:~~g_1=0.212,~~~g_2=0.84\ ,~~~\text{total}~~\chi^2=168 \ ,
\eea
where the first line is for the standard CSS scheme fit~\cite{Su:2014wpa}, the second for the 
TMD-scheme JMY and JCC with coefficients in Eqs.~(\ref{univerc},\ref{eq:cf1},\ref{tmdh},\ref{tmdbc}). There
is hardly any difference between the two fits. There is no difference
in the comparisons to the experimental data  either. 
It demonstrates the effective equivalence between all schemes in the phenomenological
studies. Theoretically, the difference could come from higher orders, such
as $\alpha_s^2$ in $H_{\rm TMD}$ and the coefficients at N${}^3$LL for the 
resummation which are beyond what we have considered in this paper 
and  Ref.~\cite{Su:2014wpa}.

\begin{figure}[tbp]
\centering
\includegraphics[width=5.3cm]{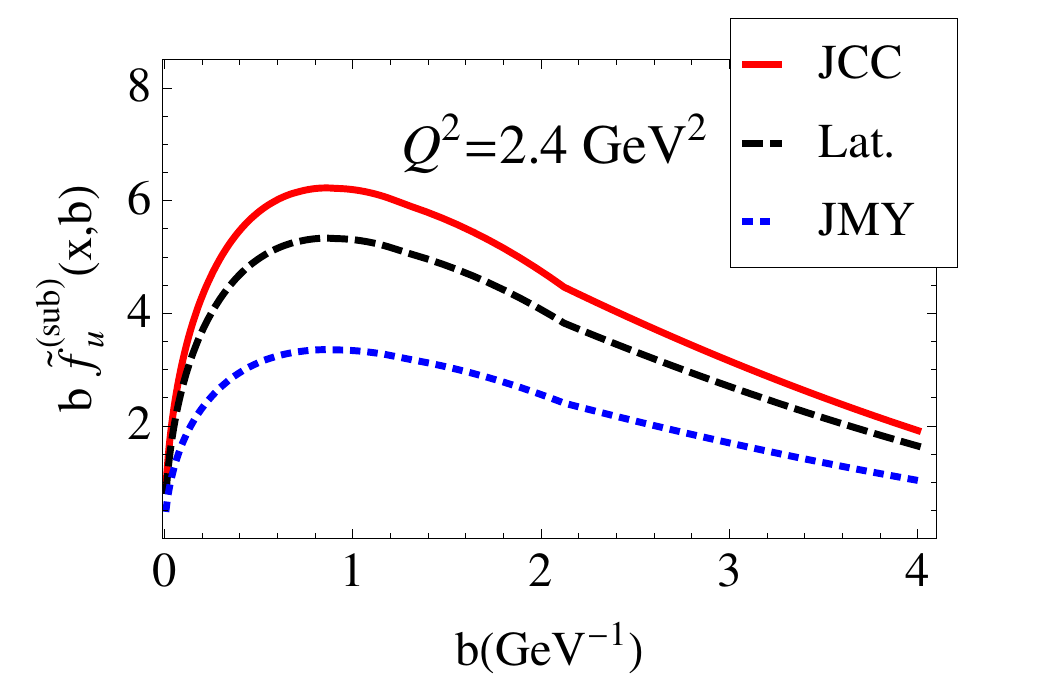}
\includegraphics[width=5.3cm]{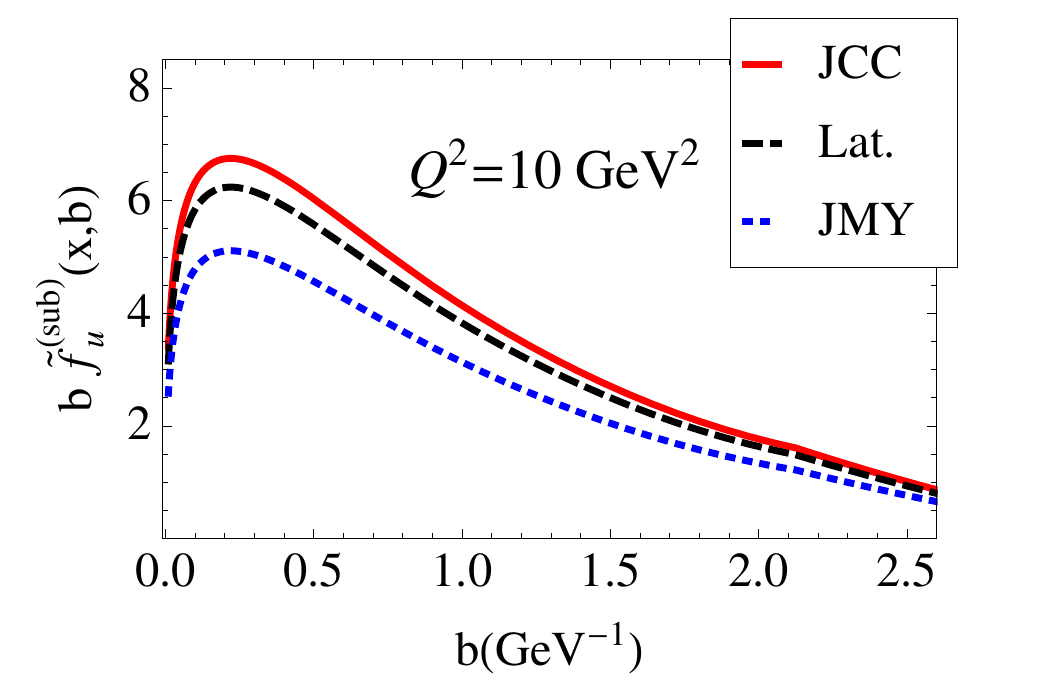}
 \includegraphics[width=5.3cm]{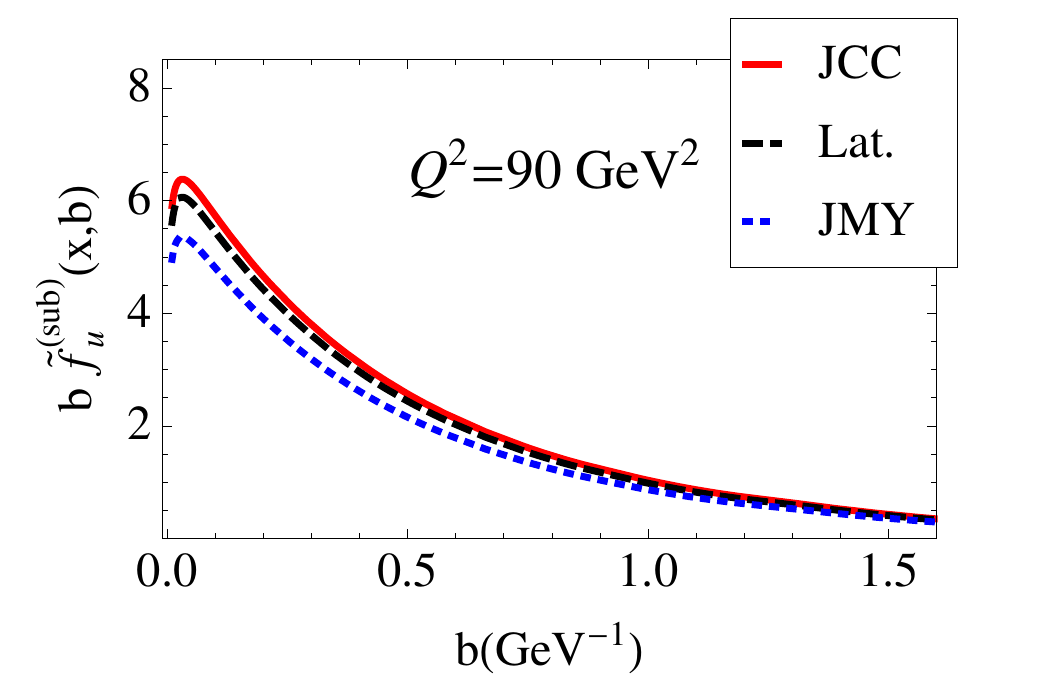}
\caption{TMD up-quark distributions $\tilde{f}_u^{(sub.)}(x=0.1,b)$
as functions of $b$  at different scale $Q^2 = 2.4$, $10$, $90$ (GeV$^2$) for 
three different schemes, from the top to the bottom: JCC~\cite{Collins:2011zzd}, Lattice~\cite{Ji:2014hxa}, JMY~\cite{Ji:2004wu,Ji:2004xq} ($\ln\rho=1$).}
\label{tmdquark_b}
\end{figure}
 
\begin{figure}[tbp]
\centering
\includegraphics[width=12.cm]{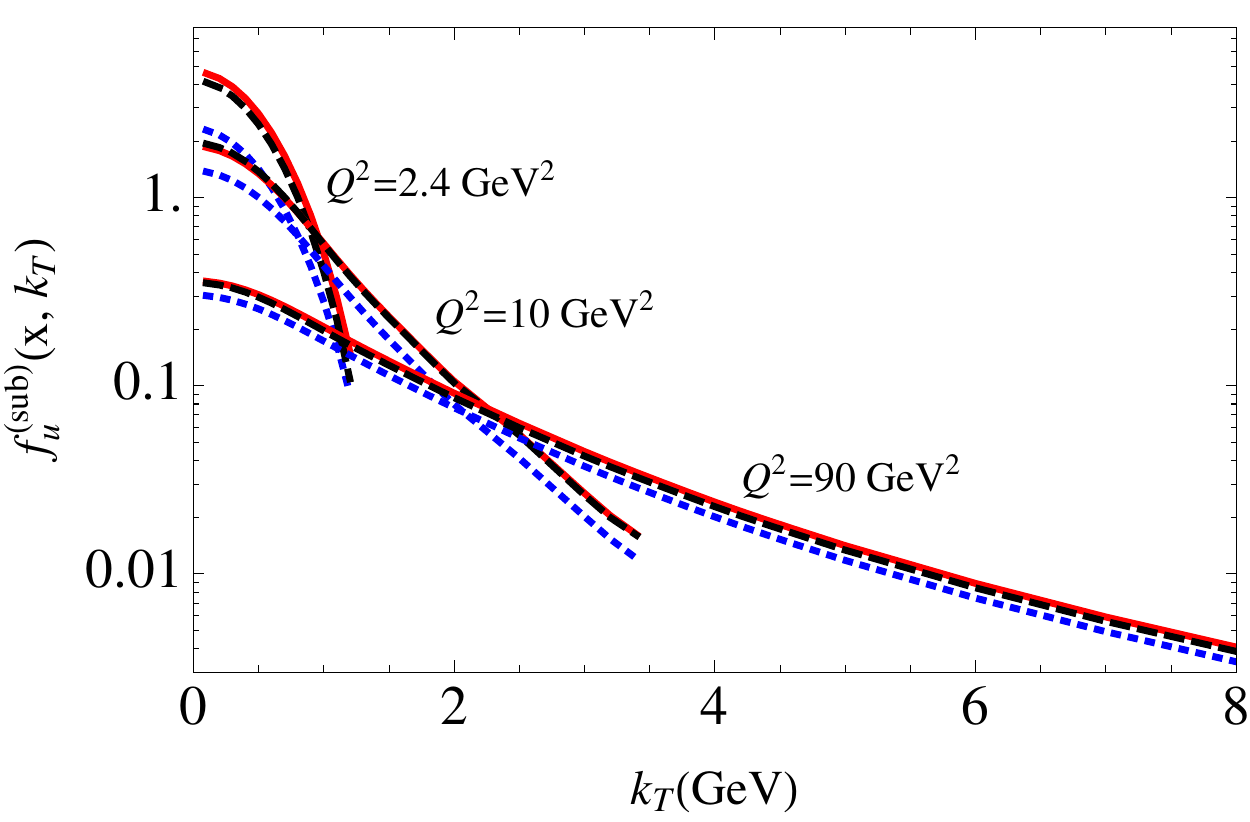}
\caption{TMD up-quark distributions  ${f}_u^{(sub.)}(x=0.1,k_\perp)$
as functions of the transverse momentum $k_\perp$ (GeV) at three different scales $Q^2 = 2.4$, $10$, $90$ (GeV$^2$) for 
three different schemes, from the top to the bottom: JCC~\cite{Collins:2011zzd}, Lattice~\cite{Ji:2014hxa}, JMY~\cite{Ji:2004wu,Ji:2004xq} ($\ln\rho=1$).}
\label{tmdquark_kt}
\end{figure}

With the non-perturbative form factors determined from the 
experimental data, we can compare the TMD quark distributions
in different schemes by evaluating them using Eqs.~(\ref{jccf},~\ref{latf},~\ref{tmdqf}). Ji-Ma-Yuan 2004 scheme has a residual dependence on the 
value of $\rho$, we fix it by choosing $\ln \rho =1$.The transverse momentum dependence in three schemes 
is calculated by Fourier transformation respect
to $b$ using Eqs.~(\ref{tmdqf},\ref{tmdqfc},\ref{tmdqflat}). In Fig.~\ref{tmdquark_b}, as an example,
we plot the up-quark distributions $b \tilde{f}_{u}^{(sub)}(x=0.1,b)$ at $x=0.1$ for 
different schemes at different  scale $Q^2 = 2.4$, $10$, $90$ (GeV$^2$) 
as functions of $b$ (GeV$^{-1}$) and in Fig.~\ref{tmdquark_kt} we plot $f_u^{(sub)}(x=0.1,k_\perp)$ as 
function of the transverse momentum $k_\perp$.   
One can see from Fig.~\ref{tmdquark_b} that at low values of $Q^2$ the 
non-perturbative part of the distribution becomes very important
and the values of $b>b_{max}$ dominate the result in $k_\perp$ space. 
At higher values of $Q^2$ the large $b$ tail of the distribution is suppressed 
and the whole distribution can be computed using mainly perturbative regime $b < b_{max} $.
In this regime the results will have a relatively low sensitivity to the 
non-perturbative input of TMD evolution. Again, the difference
between different schemes is due to the coefficient
$\widetilde{\cal F}_q$ in Eqs.~(\ref{jmyf},\ref{jccf},\ref{latf}). Because
the difference is proportional to $\alpha_s(Q)$, it will become 
smaller at higher scale $Q$ as shown in Figs.~\ref{tmdquark_b},~\ref{tmdquark_kt}.
Similar plots have been shown in Ref.~\cite{Aybat:2011zv} for the quark distributions
in the JCC scheme, however, using the previous BLNY parameterization~\cite{Landry:2002ix}
for the non-perturbative form factors obtained in CSS resummation. 
In our calculation we have consistently used relation between different schemes and the fitted non-perturbative 
form factors.

\section{Conclusion\label{Conclusions}}

In this paper, we have investigated the scheme dependence in the
TMD parton distributions and factorizations to describe the experimental data 
of hard scattering processes in hadron collisions. The equivalence between
different schemes   can be proven in perturbation theory order by order 
following the procedure of a similar study of Catani-de Florian-Grazzini  
2000~\cite{Catani:2000vq}. We have 
studied three such schemes, Collins 2011~\cite{Collins:2011zzd}, 
Lattice~\cite{Ji:2014hxa}, or Ji-Ma-Yuan 2004~\cite{Ji:2004wu,Ji:2004xq}, and 
have demonstrated the equivalence between them and equivalence to the 
standard CSS method. The associated coefficients are illustrated at one-loop order.

With TMD scheme dependence embedded in the coefficients, $\widetilde{\cal F}$ and $ 
{\cal H}^{\rm TMD}$, the resummation formulas 
have been applied to the Drell-Yan type of lepton pair production in $pp$
collisions, and the associated non-perturbative form factors are determined
from the global fit. We found that the TMD-schemes produce the same
phenomenological results as compared to the standard CSS scheme
for the resummation. More importantly, the parameters of the associated non-perturbative
form factors are also found to be the same in all schemes. 

Using the fitted parameters, we can calculate the TMD quark distributions
as functions of the transverse momentum. We have compared
the results from three different schemes. This comparison becomes useful
in the TMD interpretation of the experimental results.

In this paper we explored  the spin-average quark distributions. Similar
studies can be carried out for all other TMDs. In particular, the
quark Sivers function, which describes the correlation of the transverse 
momentum of the quark and the nucleon spin, can be formulated in
the CSS resummation formalism. The non-perturbative form factors for 
Sivers function, however,
will be different from the unpolarized quark distributions discussed
in this paper. In order to determine these non-perturbative factors one 
needs to perform a global fit to the existing 
SIDIS data. We leave that for a future publication.

\section{Acknowledgments}
We thank C.~P.~Yuan for discussions and comments. 
This material is based upon work supported by the U.S. Department of Energy, 
Office of Science, Office of Nuclear Physics, under contracts No.~
DE-AC02-05CH11231 (PS, FY), No.~DE-AC05-06OR23177 (AP).

\section*{References}

\bibliographystyle{elsarticle-num}
\bibliography{collins-biblio}

\end{document}